\documentstyle[11pt]{article}

\def\be{\begin{equation}}
\def\ee{\end{equation}}

\begin{document} \thispagestyle{empty} \begin{flushright}
\framebox{\small BRX-TH~535}\\
\end{flushright}

\vspace{.8cm} \setcounter{footnote}{0}

\begin{center} {\Large{\bf  Stress and Strain: $\mbox{\boldmath
 $T^{\mu\nu}$}$ of Higher Spin Gauge Fields}
    }\\[8mm]

S. Deser\\
Department of Physics, Brandeis University\\ Waltham, MA 02454,
USA

Andrew Waldron\\
Department of Mathematics, University of California\\  One Shields
Avenue, Davis, CA 95616, USA

{\small (\today)}\\[1cm]
\end{center}

\begin{quotation}
Abstract: We present some results concerning local currents,
particularly the stress tensors $T^{\mu\nu}$, of free higher
($>$1) spin gauge fields. While the $T^{\mu\nu}$ are known to be
gauge variant, we can express them, at the cost of manifest
Lorentz invariance, solely in terms of (spatially nonlocal)
gauge-invariant field components, where the ``scalar" and ``spin"
aspects of the systems can be clearly separated.  Using the
fundamental commutators of these transverse-traceless variables we
verify the Poincar\'{e} algebra among its generators, constructed
from the $T^0_\mu$ and their moments.  The relevance to the
interaction difficulties of higher spin systems is mentioned.
\end{quotation}

  \section{Introduction}

Gauge fields with spin $>$1 enjoy a (deservedly) bad reputation,
at least when they are not parts of some infinite tower, possibly
in nonflat backgrounds.  Their best-known problems lie in the
difficulty of consistent interaction with gravity, and (with the
sole, but significant, exception of self-interaction spin 2, {\it
aka} general relativity) with themselves.  Here we give a brief
and preliminary discussion of work  \cite{001} of a related but
rather different aspect of gauge field problems, namely the
unavoidable gauge variance of their stress-tensors $T^{\mu\nu} =
T^{\nu\mu}$; non-symmetric tensors are uninteresting because they
do not define rotation generators. What has been known for some
time \cite{002} is that those of their spatially integrated
moments corresponding to the Poincar\'{e} generators $(P_\mu ,
J^{\alpha\beta})$ are nevertheless both gauge invariant and obey
the Poincar\'{e} algebra ensuring the invariance of the underlying
theory.  More precisely, the result of \cite{002} is that all the
gauge dependence of the $T^{0\mu}$ is concentrated in identically
conserved superpotential $S^{0\mu} \equiv
\partial^2_{\alpha\beta} \Delta^{0\alpha\mu\beta}$, where the
$\Delta$ have the algebraic symmetries of the Riemann tensor.
This means that $S^{00} \equiv
\partial^2_{ij} \: \Delta^{0i0j}$ can contribute neither to the
energy $\int d^3x \, T^0_0$, nor to the boosts $K^i =
 \int d^3x \; x^ i T^{00}$.  Likewise, the effect of
$S^{0i} \equiv \partial^2_{k\ell} \Delta^{0ki\ell} +
\partial^2_{k0} \: \Delta^{0ki0}$ vanishes both in the momentum
$\int d^3x \: T^0_i$ and in the rotation generators $\int d^3x
(x^i T^{0j} - x^j T^{0i})$.  [That the first part of $S^{0i}$ does
not contribute is obvious; the second doesn't due to the $(ik)$
symmetry of $\Delta^{0ki0}$.]  So all free fields in flat space
are indeed safe (we will also discuss deSitter backgrounds in
\cite{001}).  In fact one could finish the discussion right here
by noting that as long as a current is conserved on shell in a
field theory, and plays no dynamical role as a ``source" then all
that need be demanded of it is that its spatial integrals produce
gauge invariant generators of the corresponding transformation
(here Poincar\'{e} rotations).  But then free fields are always
well-behaved and dull.  It is in their role and effects when
interaction is introduced that the local currents must come under
scrutiny; at the very least they must consistently couple to
gravity, so their  $T^{\mu\nu}$ do count!

That  $T^{\mu\nu}$ is gauge variant for spin $>$1 (we stick to
massless, gauge, fields throughout) is obvious:  The form of
$T^{\mu\nu}$, in terms of potentials, $\phi_{\alpha\beta \ldots}$
is of course $T \sim a \: \partial\phi \: \partial\phi + b \:
\phi\partial^2 \phi$;  the $b$ term can generally be exchanged for
a superpotential.  [Correspondingly $T \sim \psi
 \partial \psi$ for fermions, which face the same problems for
 $s>\frac{1}{2}$, except
 that there is one less derivative.  We
 leave this parallel story to the reader.]  But for $s>1$ at least
 two derivatives are required to define local gauge invariants,
 namely (linearized) ``curvatures". [The Maxwell tensor is safe
 because the curl of $A_\mu$ is already invariant, and
 $A\partial^2A$ terms needn't appear.]  From this point of view,
 it seems miraculous that the Poincar\'{e} generators are
 invariant; our approach should also dispel this paradox.

\section{Vector Currents}

It is perhaps instructive to note first that loss of gauge
invariance in currents already occurs at spin 1, where the current
is the vector $j^\mu$ associated with invariance under internal
rotations of a multiplet of vector gauge fields, rather than with
the space-time invariances of $T^{\mu\nu}$.  The simplest example
is a doublet, the complex field $C_\mu$ obeying Maxwell's
equations. The associated (neutral) current is $j^\mu =
(iG^{\mu\nu *} C_\nu + c.c.)$, unique up to superpotentials
$\Delta j^\mu \equiv
\partial_\nu \Sigma^{\mu\nu}$,  $\Sigma^{\mu\nu} = -
\Sigma^{\nu\mu}$.  The dependence on the potential $C_\nu$, and
not only on the field strength $G_{\mu\nu} = \partial_\mu C_\nu -
\partial_\nu C_\mu$, is the source of gauge variance here, and
cannot be ``improved" away.  The total charge $Q = \int  d^3x \,
j^0$ is (of course) conserved and gauge invariant (on shell) under
the local gauge transformation: since $j^0 \sim G^{0i*} C_i$, then
$\delta j^0 \rightarrow \partial_i (G^{0i*} \, \Lambda )$ under
$\delta C_i = \delta_i \, \Lambda$, owing to the Gauss constraint
$\partial_i \, G^{0i} = 0$.  The ``gauge-invariant" form of $j^0$
is best exhibited in radiation gauge, where $A_i$, like $G^{0i}$,
is also transverse: $j_0 \sim G^{*T}_{0i} \, A^T_i$, or
equivalently $j_0 \sim G^T_{0i} \, A^T_i + \partial_i (G^T_{0i} \,
A^L )$.  The analogy to stress tensors can be taken one more step.
Just as spin 2 can be deformed to GR when coupled to its stress
tensor, so does the coupling of complex field above to $j^\mu$
deform to become Yang--Mills when a third, neutral, $A_\mu$ field
is introduced to provide the $j^\mu (C) A_\mu$ interaction
\cite{003} and complete the triplet, $(C_\mu , A_\mu )$.

\section{Stress Tensors}

\renewcommand{\theequation}{3.\arabic{equation}}
\setcounter{equation}{0}

Our aim is to exploit the fact that renouncing manifest Lorentz
invariance allows us to restate $T^{\mu\nu}$ in ``manifestly"
gauge-invariant form -- making gauge choices to be sure.

The existence of gauge invariant representations of $T^{\mu\nu}$
can be understood as follows.  For concreteness, we use $D$=4,
where all $s > 0$ gauge fields have just two independent, helicity
$\pm s$, modes, obeying the wave equation.  [These theories are
formulated in terms of totally symmetric tensors
$\varphi_{\mu_1\ldots\mu_s}$ (subject to double tracelessness for
$s>3$).]  They correspond (say in symmetric tensor representation)
to (spatially nonlocal) transverse-traceless spatial component
$\phi^{TT}_{ij\ldots}$. The other field components are constraint
variables, Lagrange multipliers or pure gauges.  Hence the
original, gauge invariant, action\footnote{In generic
gravitational backgrounds, this invariance is lost and is a
symptom of the gravitation interaction problems.} will reduce,
upon enforcing the constraints, {\it but in any gauge,} to the
simple ``oscillator" form
\begin{equation}
I = \int d^4x \; \left[ \sum^2_1 p_A\dot{q}_A - H (p,q) \right] \;
, \;\;\;\; H = \frac{1}{2} \; \Big\{ p^2 + (\nabla q )^2 \Big\} \;
.
 \end{equation}
 The two conjugate pairs $(p_A , q_A)$ denote the appropriate
 spatial $TT$ variables, with implicit summation over indices.
 While this representation of the action seems to contain just two
``scalars", the tensorial nature of the variables is implicit in
their $TT$ nature.  Hence the generators must be of the form
\begin{eqnarray}
P_0 & = & \int d^3x \: H \; ,\;\;\;\; \mbox{\boldmath{$P$}} =
-\int \, d^3x
\: p \mbox{\boldmath{$\nabla$}}q \nonumber \\
\mbox{\boldmath{$J$}} & = & \int d^3x \: \{
p(\mbox{\boldmath{$r$}}\times \mbox{\boldmath{$\nabla$}})q + s \:
p\times q \}
\: = \:\mbox{\boldmath{$L$}} + \mbox{\boldmath{$S$}} \nonumber \\
\mbox{\boldmath{$K$}} & = & \int d^3x \: \mbox{\boldmath{$r$}} H
-t \mbox{\boldmath{$P$}}
\end{eqnarray}
where the spin term $\mbox{\boldmath{$S$}}$ is shorthand for a
suitable index contraction scheme (see below). [These generators
obey the Poincar\'e algebra at any time $t$, in particular one can
set $t=0$ in ${\bf K}$. However, since $[\mbox{\boldmath{$K$}}
,P_0] \neq 0$ --boosts are time dependent-- only ${\bf K}(t)$
generates symmetry transformations of the action (3.1).]   This
``prediction" in turn implies the existence of a set, $T^{\mu\nu}
(p,q) = T^{\nu\mu}$, that yields the moments (3.2), and obeys
on-shell conservation, $\partial_\mu T^{\mu\nu} = 0$, where
$p=\dot{q}$, $\Box q = 0$. Indeed, we can be even more explicit
and predict that
\begin{equation}
T^{00} = H + S^{00} \; , \;\;\;\; T^0_i = - p \partial_i q + s
\partial_j (p^{j\ell\ldots}q_{i\ell\ldots}) + S^0_i
\end{equation}
where $S^{0\mu}$ are superpotentials.  The only role of $T_{ij}$
is to verify that $\partial_0 \, T^0_i$ is a spatial divergence,
and that is in turn guaranteed by the form (3.3).  Furthermore the
fundamental commutation relations,
\begin{equation}
i[p_A,q^\prime_B ] = [ \delta_{AB} ( \mbox{\boldmath{$r$}} -
\mbox{\boldmath{$r$}}^\prime ) ]^{TT} \; ,
\end{equation}
where the right side is $TT$ in each of its variables, guarantee
(but non-trivially as we shall note) the Poincar\'{e} algebra
among the integrated generators (3.2).

\section{Spin 1}

\renewcommand{\theequation}{4.\arabic{equation}}
\setcounter{equation}{0}

We now illustrate the above requirements first with a parallel
treatment of the Maxwell field before going on to the main, $s>1$,
case.  The general procedure, after fixing on a candidate
conserved $T^{\mu\nu}$  (all of which differ by a superpotential),
is to insert the constraints, remove their Legrange multipliers
and fix gauge variables, just leaving the desired gauge invariant
pairs $(p_A , q_A )$.  [All such gauge fixings are also only a
superpotential away from each other.]

Let's begin with the ``degenerate" spin 1 case, where no gauge
fixing is needed, but constraints still must be solved, as part of
the ``on-shell" procedure.  In first order form, where $(-
\mbox{\boldmath{$E$}}_T , \, \mbox{\boldmath{$A$}}^T)$ are the
transverse conjugate variables and $\mbox{\boldmath{$B$}} =
\mbox{\boldmath{$\nabla$}} \times \mbox{\boldmath{$A$}}^T$,
\begin{eqnarray}
T^{00}& = & \textstyle{\frac{1}{2}} \: (\mbox{\boldmath{$E$}}^2
+\mbox{\boldmath{$B$}}^2)  =  \textstyle{\frac{1}{2}} \: (p^2 +
(\nabla q )^2 ) - \textstyle{\frac{1}{2}} \: \partial^2_{ij}
(q_iq_j) \equiv H_s + \partial^2_{ij} \Delta^{ij}
\nonumber \\
T^0_i & = & (E \times B)_i  =  - p\partial_i \, q + \partial_j
(p,q_i)  \equiv  T^0_s{_i} + \partial_j (p_jq_i) \; .
\end{eqnarray}
Note the spin term in $T^0_i$, with unit coefficient, and the fact
that apart from it, the rest of the $T^0_\mu$ are of ``scalar",
$H_s = \frac{1}{2} (p^2 + (\nabla q)^2 ), \; T^0_s{_i} =
-p\partial_i \, q$, form.  There is an important lesson here: The
Maxwell Poincar\'{e} generators are (taking the boosts at $t=0$
for simplicity)
\begin{equation}
P_\mu = P_\mu (s) \; , \;\;\;\; K^i = \int d^3x \: x^i T^{00} =
K^i(s) \; , \;\;\;\;\mbox{\boldmath{$J$}} =
\mbox{\boldmath{$L$}}(s) + \int d^3x \, (p \times q) \; .
\end{equation}
Apart from the extra spin term in (4.2) they are of pure scalar
form.  But since the scalar generators ``certainly" satisfy the
scalar algebra, how does the spin term ever appear in the
boost-boost commutator, $[K^i, \, K^j] = \epsilon^{ijk}J_k$, since
$[K^i (s),K^j(s)] = \epsilon^{ijk} L_k(s)$ only?  The answer is
subtle and makes essential use of the fact that our variables are
transverse, hence obey the transverse fundamental equal-time
commutator,
\begin{equation}
i [p_i,q_{j{^\prime} }] = [\delta_{ij{^\prime}}
(\mbox{\boldmath{$r$}}- \mbox{\boldmath{$r$}}^\prime
)]^{TT{^\prime}} \equiv
\delta_{ij{^\prime}}\delta^3(\mbox{\boldmath{$r$}}-\mbox{\boldmath{$r$}}^\prime
)  + \partial_i\partial_{j{^\prime}}
G(\mbox{\boldmath{$r$}}-\mbox{\boldmath{$r$}}^\prime)
\end{equation}
where $G$ is the Coulomb Green Function.  If one keeps track of
this extra term, then the spin part $\mbox{\boldmath{$S$}}$ duly
appears.  The lesson is that whenever moments are involved, it is
essential to operate in the correct space of transverse
(-traceless) tensors.

\section{Spin $\mbox{\boldmath{$\geq$}}$ 2}

\renewcommand{\theequation}{5.\arabic{equation}}
\setcounter{equation}{0}

Let us (at last) turn to the first system of interest, spin 2,
{\it i.e.}, the linearized approximation of GR about (say) flat
space (all higher spins behave the same way).  There are (as we
know from GR) infinitely many candidate $T^{\mu\nu}$ differing by
superpotentials, and none is (abelian) gauge invariant.  One
example is the Landau--Lifshitz complex, which is long but
involves only bilinears in the (linearized)
$\Gamma_{\mu\nu}^\alpha$; of course it yields the same
$T^{\mu\nu}$ as our choice below, up to a superpotential.  More
useful is simply the quadratic part of the Einstein tensor, which
we adopt here; one advantage of choosing it,
\begin{equation}
T_{\mu\nu} \equiv - \textstyle{\frac{1}{4}} \: G^Q_{\mu\nu}
\end{equation}
is that its conservation is an immediate consequence of the
Bianchi identities (at each order in a field expansion) and the
on-shell conditions, $G^L_{\mu\nu} = 0$.  [The suffixes $(Q,L)$
stand for (quadratic, linear) expression in $h_{\mu\nu} \equiv
g_{\mu\nu} - \eta_{\mu\nu}$].  Indeed,
\begin{equation}
0 \equiv (D_\mu G^{\mu\nu})_Q \equiv \partial_\mu G^{\mu\nu}_Q +
(\Gamma^L G_L )^\nu = \partial_\mu G^{\mu\nu}_Q \; .
\end{equation}
The next advantage, aside from not having to exhibit $T^{ij}_Q$
(since we know there is one!), is that the $G^{0\mu}$ are already
the energy-momentum constraints in the full theory.  Even a manual
calculation is not too difficult, in terms of the only remaining
variables $(\dot{h}^{TT}_{ij} ,h^{TT}_{ij}) \equiv$ $(p_A,q_A), \;
A = 1,2$ after constraints and gauge choices are imposed.

One finds the same story as for spin 1,
\begin{eqnarray}
T^{00} & \cong & \textstyle{\frac{1}{2}} (p^2 + (\nabla q)^2 ) \nonumber \\
T^{0}_i & \cong & -p \partial_i q - 2 \partial_j (p_{jk}q_{ik})
\end{eqnarray}
where $\cong$ means up to superpotentials.  The fundamental
$(p_A,q_A^\prime )$ ETC are suitably higher-index
transverse-traceless versions of the vector case and one may
easily extend (5.3) to arbitrary spins using suitable notation to
generalize ``$p\times q$".  The integrated generators are of the
``mostly scalar" Maxwell form as well, except for the ``bigger"
spin term $\mbox{\boldmath{$S$}}$.

\section{Conclusion}

\renewcommand{\theequation}{6.\arabic{equation}}
\setcounter{equation}{0}

We begin with some additional comments:  (1) There exist, of
course, various representations of the field variables, such as
vierbein form involving a non-symmetric $e_{\mu\alpha}$ and a
connection $\omega^{\alpha\beta}_\mu$.  The action can even be
made to resemble a multi-photon system with ``internal" index
$\alpha$. But a spin 2 system is {\it not} merely a photon
multiplet, a difference that ruins the gauge invariance of the
associated symmetric $T^{\mu\nu}_s (e, \omega)$.  In this
formulation, the canonical $T^{\mu\nu}_c$ does retain gauge
invariance, but to no avail: only symmetric $T^{\mu\nu}$ can
define angular momentum. More generally, field redefinitions
cannot cure the basic noninvariance problem for $s=2$, nor {\it a
fortiori} for higher spins.  (2) The formulation we have employed
here can perhaps be generalized to constant curvature backgrounds
\cite{001} but not, as noted, to generic curved spaces.  (3) The
conditions for Lorentz invariance were realized here by computing
the commutation relations among the Poincar\'{e} generators. There
is also a well-known local criterion for Lorentz invariance in
QFT, namely the Dirac-Schwinger ETC.
\begin{equation}
i[T_{00} (\mbox{\boldmath{$r$}}), \; T_{00}
(\mbox{\boldmath{$r$}}^\prime )] = (T^{0i}\partial_i +
T^{0i{^\prime}}\partial_i^\prime ) \delta^3
(\mbox{\boldmath{$r$}}-\mbox{\boldmath{$r$}}^\prime ) \; .
\end{equation}
This form is however, inapplicable here because of the
non-manifestly covariant form of our $T^0_\mu$.  For example
adding Lorentz-variant terms to $T^{\mu\nu}$ is always detected in
the integrated ETC, but not in the local form (5.4).  This is a
difference worth pursuing.

Finally, we emphasize that our new stress tensors, like the
covariant ones, are still not suitable for coupling, as currents,
to gravity because that requires both invariances to be
simultaneously manifest.  Indeed, were there a spin 2 gauge
invariant tensor, it would imply existence of a consistent cubic
self-interacting model of gravity, with abelian invariance and
without higher derivatives. Finding consistent dynamical sources
of generic $s>1$ fields seems even more unlikely in any local
context.

This work was supported by the National Science Foundation under
grants PHY99-73935 and PHY01-40365.


\begin{thebibliography}{999}
  \bibitem{001}
  S.\ Deser and A.\ Waldron, in preparation.
  \bibitem{002}
  S.\ Deser and J.M.\ McCarthy, Class.\ Quant.\ Grav.\ {\bf 7} (1990) L119.
  \bibitem{003}
  R.A.\ Arnowitt and S.\ Deser, Nucl.\ Phys.\ {\bf 49} (1963) 163.
 \end{thebibliography}
\end{document}